\documentclass[10pt, conference]{IEEEtran}
\IEEEoverridecommandlockouts
\usepackage{cite}
\usepackage{booktabs}
\usepackage{amsmath,amssymb,amsfonts}
\usepackage{algorithmic}
\usepackage{graphicx}
\usepackage{threeparttable}
\usepackage[multiple]{footmisc}
\usepackage{textcomp}
\usepackage{xcolor}
\def\BibTeX{{\rm B\kern-.05em{\sc i\kern-.025em b}\kern-.08em
    T\kern-.1667em\lower.7ex\hbox{E}\kern-.125emX}}
\begin{document}

\title{ManyTypes4Py: A Benchmark Python Dataset for Machine Learning-based Type Inference\\
}

\author{\IEEEauthorblockN{Amir M. Mir}
\IEEEauthorblockA{\textit{Department of Software Technology} \\
\textit{Delft University of Technology}\\
Delft, The Netherlands \\
s.a.m.mir@tudelft.nl}
\and
\IEEEauthorblockN{Evaldas Latoškinas}
\IEEEauthorblockA{\textit{Department of Software Technology} \\
\textit{Delft University of Technology}\\
Delft, The Netherlands \\
e.latoskinas@student.tudelft.nl}
\and
\IEEEauthorblockN{Georgios Gousios}
\IEEEauthorblockA{\textit{Department of Software Technology} \\
\textit{Delft University of Technology}\\
Delft, The Netherlands \\
g.gousios@tudelft.nl}
}

\maketitle

\begin{abstract}
In this paper, we present ManyTypes4Py, a large Python dataset for machine learning (ML)-based type inference. The dataset contains a total of 5,382 Python projects with more than 869K type annotations. Duplicate source code files were removed to eliminate the negative effect of the duplication bias. To facilitate training and evaluation of ML models, the dataset was split into training, validation and test sets by files. To extract type information from abstract syntax trees (ASTs), a light-weight static analyzer pipeline is developed and accompanied with the dataset. Using this pipeline, the collected Python projects were analyzed and the results of the AST analysis were stored in JSON-formatted files. The ManyTypes4Py dataset is shared on zenodo and its tools are publicly available on GitHub.
\end{abstract}


\section{Introduction}
In recent years, dynamic programming languages (DPLs) have become immensely popular as they give developers fast prototyping \cite{ieeespec2019}. However, DPLs lack static typing, which causes several issues such as unexpected run-time exceptions, sub-optimal support for integrated development environments (IDEs), and less precise program analysis. To address these issues, optional static typing is introduced for DPLs like Python \cite{van2014pep}, JavaScript \cite{bierman2014understanding}, and PHP \cite{klingstrom2020type}. Yet, developers are required to manually add type annotations to their existing codebases, which is a laborious task \cite{ore2018assessing}. To ease the type annotation burden, researchers have recently employed machine learning (ML) techniques to infer types for DPLs \cite{malik2019nl2type, pradel2019typewriter, allamanis2020typilus}.

%

ML techniques need a sufficiently large dataset to achieve an acceptable level of generalization for the task at hand \cite{sun2017revisiting}. Concerning the ML-based type inference for DPLs, it is difficult to create a benchmark dataset that contains software projects with a sufficient number of type annotations. Because of the optional static typing, many software projects written in DPLs lack type annotations. Nevertheless, to train an ML-based type inference model for Python, researchers created their own dataset by either gathering a small set of projects with type annotations \cite{pradel2019typewriter} or employ static type inference tools to add type annotations to existing projects \cite{allamanis2020typilus}.

We believe that there is a need for a large benchmark dataset that facilitates training ML-based type inference models, especially for Python. Unlike TypeScript's compiler, the Python interpreter cannot infer the type of variables or function signatures at compile time \cite{scott16}. Motivated by this, we present the ManyTypes4Py dataset, a large dataset to train ML models for predicting type annotations in Python. Currently, we are working on the Type4Py model \cite{mir2021type4py}, which is trained on the earlier version of the ManyTypes4Py dataset. The experimental results show that the model trained on our dataset is overall more accurate when compared to the same model trained on a smaller dataset \cite{mir2021type4py}.

In summary, the paper has the following contributions:
\begin{itemize}
	\item \textbf{ManyTypes4Py dataset}, which features 5,382 Python projects with more than 869K type annotations. The latest version of the dataset can be downloaded on zenodo\footnote{https://zenodo.org/record/4479714}.
	\item \textbf{LibSA4Py tool}, a light-weight static analyzer pipeline to process Python projects and extract type hints/features for training ML-based type inference models. The tool is publicly available on a GitHub repository\footnote{https://github.com/saltudelft/libsa4py}.
\end{itemize}

\begin{figure*}[!t]
	\centering
	\includegraphics[width=0.8\linewidth]{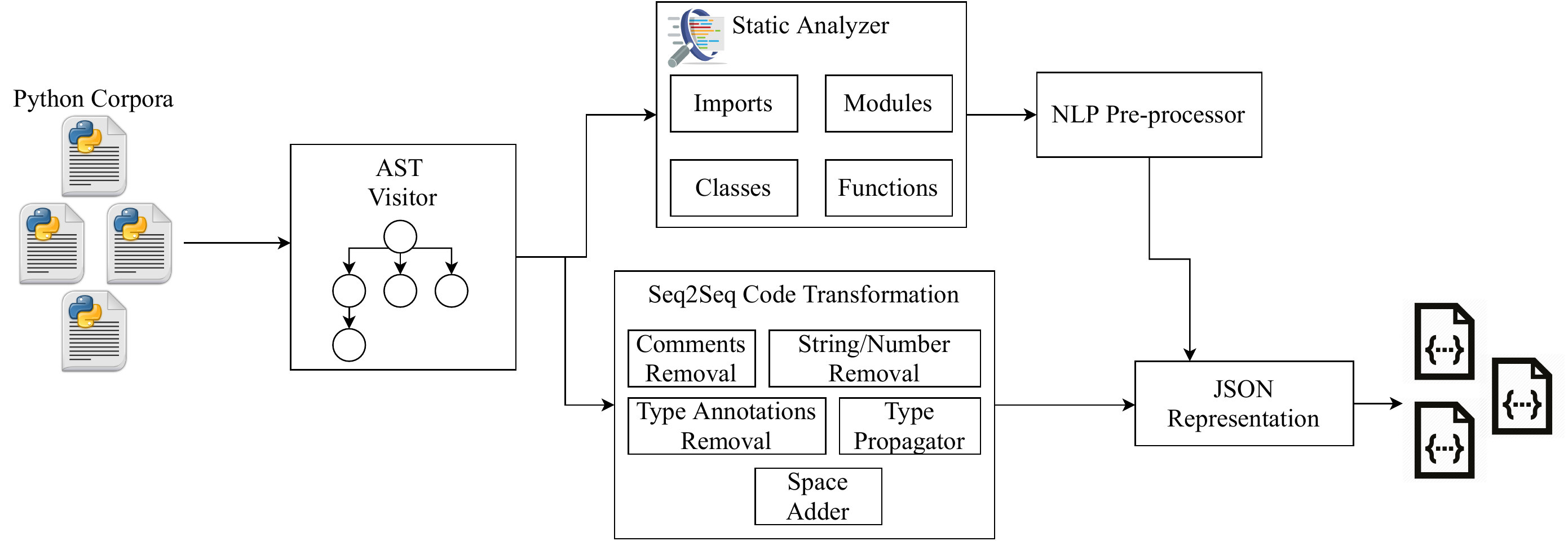}
	\caption{Overview of light-weight static analysis pipeline (LibSA4Py tool)}
	\label{fig:overview-pipeline-sa}
\end{figure*}

\section{Method}\label{sec:method}
We created the ManyType4Py dataset using the following methodology:

\begin{itemize}
	\item To find Python projects with type annotations, we intuitively search for projects that have mypy as a depdendency on libraries.io. Since mypy is the most used type checker for Python, projects that use mypy have most likely type annotations. Our search resulted in 5,382 Python projects that are publicly available on GitHub. We cloned all the discovered projects in Sep. 2020 and created a file that contains projects' URL and their latest commit hash.
	\item As demonstrated by Allamanis \cite{allamanis2019adverse}, it is essential to de-duplicate a code corpora before training ML models, as code duplication negatively affects the performance of ML models when testing on duplicated code corpora. Following this, we de-duplicated the collected Python corpora using our code de-duplication tool, namely, CD4Py\footnote{https://github.com/saltudelft/CD4Py}. In short, the CD4Py tool tokenizes Python source code files, vectorizes files using Term Frequency-Inverse Document Frequency (TF-IDF), and performs $k$-nearest neighbor search to identify candidate duplicates files.
	\item After removing duplicate files, we split the dataset into three sets by files, i.e., 70\% training data, 10\% validation data, and 20\% test data. This is a common practice that is considered in recent research work \cite{pradel2019typewriter, allamanis2020typilus}, concerning machine learning-based models for type inference.
	\item Given the de-duplicated code corpora and a list of files for the three sets, we ran light-weight static analysis pipeline, which is depicted in Figure \ref{fig:overview-pipeline-sa}. First, the Abstract Syntax Tree (AST) of Python source files are visited. Second, type hints and features are statically extracted from imports, modules, classes, and functions, inspired by recent ML-based type inference approaches \cite{pradel2019typewriter, malik2019nl2type}. Third, the seq2seq representation\footnote{Each token is aligned with a type if present. Otherwise, zero is inserted.} of source code files \cite{hellendoorn2018deep} are generated by removing comments, string, number literals, and propagating types. Forth, common Natural Language Processing (NLP) practices such as tokenization and lemmatization are applied to identifier names in source code files. Finally, the processed Python projects are stored as a JSON-formatted file. 
\end{itemize}

After completing all the aforementioned steps, a zip file is created which contains: (1) JSON file of processed Python projects (2) a file containing projects' URL and their latest commit hash (3) a file containing duplicate files in the dataset (4) a CSV file containing a list of files and their corresponding set. The helper scripts and instructions for preparing the dataset are publicly available on a GitHub repository\footnote{https://github.com/saltudelft/many-types-4-py-dataset}.

\begin{table}[!t]
	\centering
	\caption{Duplication statistics across the ManyTypes4Py dataset}
	\label{tab:dup-stats}
	\begin{tabular}{l c}
		\toprule
		Duplication stats & Value \\
		\midrule
		\# Detected duplicate files & 400,245 (78.43\%) \\
		\# Detected clusters & 45,836 \\
		Avg. \# of files per clusters & 8.73 \\
		Median \# of files per clones & 3.00 \\
		Duplication ratio & 69.45\% \\
		\bottomrule
	\end{tabular}
\end{table}

\begin{table*}[!t]
	\centering
	\label{tab:char-dataset}
	\caption{Characteristics of the ManyTypes4Py Dataset}
	\begin{threeparttable}
		\begin{tabular}{l l l l l}
			\toprule
			Metrics & \multicolumn{4}{@{} c @{}}{Dataset}  \\
			\cmidrule{2-5}
			& All & Training & Validation & Test \\
			\midrule
			Repositories\tnote{a} & 5,382 & 4,913 & 2,789 & 3,796 \\
			Lines of code\tnote{b} & 22M & - & - & - \\
			\midrule
			Files & 183,916  & 132,409 & 14,675 & 36,832 \\
			...with type annotations & 50,838 (27.6\%) & 36,542 & 4,105  & 10,191  \\
			\midrule
			Functions & 2,096,797 & 1,509,048 & 169,519 & 418,230 \\
			...with comment & 1,129,573 (53.8\%) & 812,632 & 91,325 & 225,616 \\
			...with return type annotations & 325,532 (15.5\%) & 234,319 & 26,104 & 65,109 \\
			\midrule
			Arguments & 3,923,667 & 2,822,699 & 310,685 & 790,283 \\
			...with comment & 220,976 (5.6\%) & 159,453 & 16,924 & 44,599 \\ 
			...with type annotations & 480,793 (12.2\%) & 347,898 & 37,148 & 95,747 \\
			\midrule
			Types & 869,825 & 347,898 & 89,334 & 192,102 \\
			...unique & 67,060 & 53,614 & 13,995 & 23,572 \\
			\bottomrule
		\end{tabular}
		\begin{tablenotes}
			\item[a] {\footnotesize Note that there is an intersection among repositories in the three sets as the dataset is split by files.}
			\item[b] {\footnotesize Comments and blank lines are ignored when counting lines of code.}
		\end{tablenotes}
	\end{threeparttable}
\end{table*}

\section{Description}
Before describing the characteristics of the ManyTypes4Py dataset, we first describe the duplication statistics across the dataset, which is shown in Table \ref{tab:dup-stats}. The duplication ratio of the dataset is 69.45\% as detected by the CD4Py tool. This is in line with the findings of Lopes et al. \cite{lopes2017dejavu}, which showed that the Python ecosystem on GitHub has 71\% file-level duplicates. It should be noted that the duplication ratio is obtained using the following formula:
\begin{equation}
\frac{(\text{no. of duplicate files} -\text{ no. of detected clusters})}{\text{no. of source code files} * 100.0}
\end{equation}
After keeping a file from each duplicate cluster, we removed 354,409 duplicate files from the dataset.

The characteristics of the ManyTypes4Py are shown in Table \ref{tab:char-dataset} after code de-duplication. Overall, the dataset has 5,382 Python projects and 183,916 source code files (i.e. \texttt{.py} files). 27.6\% of source code files have type annotations, i.e., there is at least one type-annotated function in those files. Of 2,096,797 functions in the dataset, 53.8\% has comments\footnote{Note that here comments are functions' docstring in Python, which can be a one-line description or a complete description of a function.}\footnote{Our LibSA4Py tool can detect Google, reST, and NumPy docstrings.} and 15.5\% has return type annotations. However, Of 3,923,667 functions' arguments, 5.6\% have comments and 12.2\% have type annotations.

\begin{figure}[!t]
	\centering
	\includegraphics[width=\linewidth]{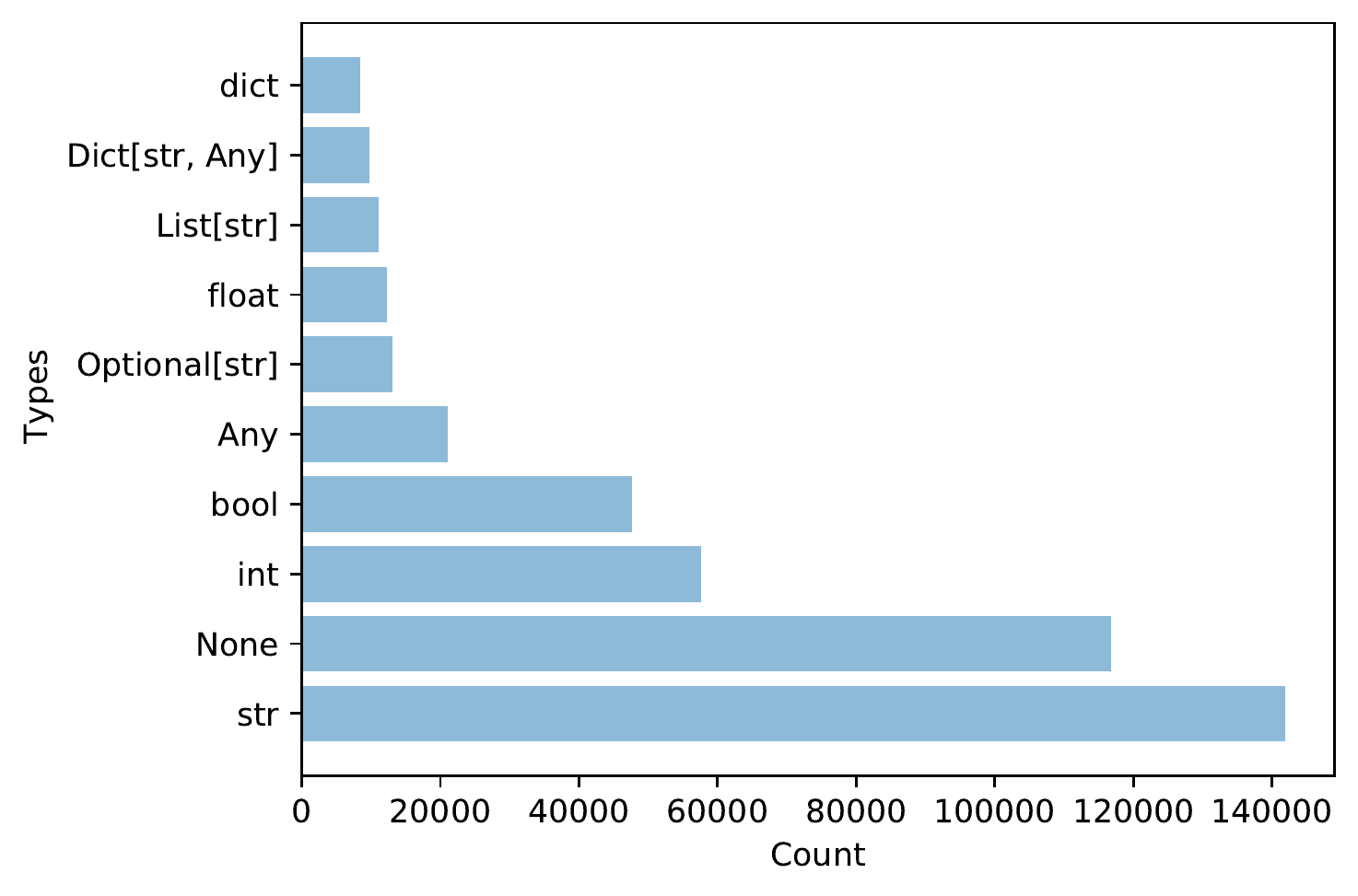}
	\caption{Top 10 most frequent types in the ManyTypes4Py dataset}
	\label{fig:top-10-types}
\end{figure}

As shown in Table \ref{tab:char-dataset}, there are a total of 869,825 type annotations and 67,060 unique types in the ManyTypes4Py dataset. To demonstrate the distribution of types, top 10 most frequent types in the dataset are shown in Figure \ref{fig:top-10-types}. Of 869,825 types, 50.56\% of them are present in the top 10 most frequent types. As can be observed from Figure \ref{fig:top-10-types}, types follow a long-tail distribution. In other words, the majority of type annotations are either \texttt{str}, \texttt{None}, \texttt{int}, or \texttt{bool}.

As stated in Section \ref{sec:method}, the dataset provides processed Python projects in JSON-formatted files, which contains various type hints and features. As of this writing, there are 23 fields in JSON-formatted files that are described in Table \ref{tab:json-fields}. Of 23 extracted fields, 16 of them are natural/contextual type hints or features that can be used for training ML-based type inference models. For instance, the \texttt{name} field stores the name of a class or a function, which is a natural source of information for predicting types \cite{malik2019nl2type}. The \texttt{ret\_exprs} and \texttt{params\_occur} fields provide return expression(s) of a function and usages of functions' parameter(s) in its body, respectively. These are considered contextual type hints, i.e., the context in which a variable or an argument is used provides a hint for predicting types \cite{pradel2019typewriter}. Also, the \texttt{untyped\_seq} and \texttt{typed\_seq} fields provide the normalized seq2seq representation of a Python source code file and the type of identifiers in the file, respectively. They both can directly be used for training an ML-based type inference model.

\begin{table*}[!t]
	\centering
	\caption{Description of fields in the JSON file of projects produced by the LibSA4Py pipeline}
	\label{tab:json-fields}
	\begin{tabular}{c l}
		\toprule
		Field Name in the JSON & Description \\
		\midrule
		\multicolumn{2}{c}{Project}  \\
		\midrule
		\texttt{author/repo} & The name of a project and its author on the GitHub URL  \\
		\midrule
		\texttt{src\_files} & Contains the path of a project's source code files \\
		\midrule
		\texttt{file\_path} & The path of a source code file to differentiate it with other files \\
		\midrule
		\multicolumn{2}{c}{Module}  \\
		\midrule
		\texttt{untyped\_seq} & The normalized seq2seq representation of an analyzed source code file \\
		\midrule
		\texttt{typed\_seq} & Contains the type of identifiers in \texttt{untyped\_seq} if present. Otherwise $0$ is inserted. \\
		\midrule
		\texttt{imports} & Contains the name of imports in an analyzed source code file \\
		\midrule
		\texttt{variables} & Contains variables' name and their type defined in a module (i.e. global variables) \\
		\midrule
		\texttt{classes} & Contains the JSON object of analyzed classes in a module which is described below \\
		\midrule
		\texttt{funcs} &  Contains the JSON object of analyzed functions in a module, which are described below \\
		\midrule
		\texttt{set} & The set to which a source code file belongs to, i.e., \texttt{train}, \texttt{valid}, \texttt{test} \\
		\midrule
		\multicolumn{2}{c}{Class} \\
		\midrule
		\texttt{name} & The name of an analyzed class in a module \\
		\midrule
		\texttt{variables} & Contains class variables' name and their type if present \\
		\midrule
		\texttt{funcs} & Contains the JSON object of analyzed functions in a class, which are described below \\
		\midrule
		\multicolumn{2}{c}{Function} \\
		\midrule
		\texttt{name} & The name of an analyzed function in either a class or a module \\
		\midrule
		\texttt{params} & Contains an analyzed function's parameter names and their type if present \\
		\midrule
		\texttt{ret\_exprs} & Contains the return expression(s) of an analyzed function \\
		\midrule
		\texttt{ret\_type} & The return type of an analyzed function if present \\
		\midrule
		\texttt{variables} & Contains local variables' name and their type in an analyzed function \\
		\midrule
		\texttt{params\_occur} & Contains parameters and their usages in the body of an analyzed function \\
		\midrule
		\texttt{docstring} & Contains docsting of an analyzed function if present, which has the below subfields \\
		\midrule
		\texttt{docstring.func} & One-line description of an analyzed function if present \\
		\midrule
		\texttt{docstring.ret} & Description of what an analyzed function returns if present \\
		\midrule
		\texttt{docstring.long\_descr} & Long description of an analyzed function if present \\
		\bottomrule
	\end{tabular}
\end{table*}

\section{Applications}
\paragraph{\textbf{ML-based type inference}} In this task, ML models are trained to predict the type of functions' arguments, return types, and variables for DPLs (e.g. Python). To do so, the AST of source code files are analyzed to extract features that give a hint for predicting types. By processing ASTs, the ManyTypes4Py dataset provides common features, i.e., natural and contextual type hints that can be employed to create code embeddings and train an ML model. Moreover, the provided seq2seq representation of source code files gives the full context around identifiers.

\paragraph{\textbf{Learning-based code completion}} In this application, an ML model is expected to predict part of a word or token for a function or a variable. For DPLs, code completion is a challenging task as there is no type information available. To overcome this, ASTs are statically analyzed while providing type information \cite{svyatkovskiy2019pythia}. The ManyTypes4Py dataset can be used as a baseline for training a code completion model as it provides partial type annotations for functions and variables. Also, our AST analysis pipeline (LibSA4Py tool) can further be extended to infer types of nodes and variables for simple cases.

\section{Limitations}
Currently, our static analysis pipeline cannot parse source code files in Python 2. Therefore, Python2-style type annotations\footnote{https://www.python.org/dev/peps/pep-0484/\#suggested-syntax-for-python-2-7-and-straddling-code} cannot be extracted. Due to this limitation, about 1\% of source code files in the dataset cannot be parsed.

\section{Related Work}
There are several Python code corpora that can be used for machine learning-based type inference. Recently, Allamanis et al. \cite{allamanis2020typilus} proposed the Typilus model, which is a graph-based neural model that predicts type annotations for Python. The Typilus model \cite{allamanis2020typilus} is accompanied by a dataset that contains 600 Python projects. Moreover, the source code files of Typilus' dataset are converted to graph representations that are only suitable for training the Typilus model. The ManyTypes4Py dataset provides JSON-formatted analyzed source code files that contains useful type hints for training various machine learning models. Raychev et al. \cite{raychev2016probabilistic} published the Python-150K dataset in 2016, which contains 8,422 Python projects. Unlike our dataset, the Python-150K dataset \cite{raychev2016probabilistic} is not collected solely for the ML-based type inference task, meaning that a large number of projects in the dataset may not have type annotations at all, especially given the time that the dataset was created. Allamanis \cite{allamanis2019adverse} showed that the Python-150K dataset suffers from code duplication despite the removal of project forks.

\section{Conclusion}
In this paper, we present the ManyTypes4Py dataset, a benchmark Python dataset for ML-based type inference. It contains 5,382 Python projects from GitHub with more than 869K type annotations. The collected Python projects were de-duplicated by removing duplicate source code files to ensure that trained ML models do not have duplication bias. Using the accompanying LibSA4Py tool, the AST of Python source code files were analyzed to provide 16 type hints plus a seq2seq representation for training ML-based type inference models. For each analyzed project, the result of the AST analysis is saved in a JSON-formatted file. Although the dataset's main application is ML-based type inference, it can be a useful baseline for learning-based code completion.

In the near future, we will extend our static analysis pipeline (LibSA4Py tool) to add more type annotations to the dataset by implementing cheap and simple type inference heuristics. Also, we will perform data and control flow analysis to create graph representation of source code files for training graph-based neural models. To include more projects with type annotations, we will consider projects that use other type checkers other than mypy.

\section*{Acknowledgment}
This research work was funded by H2020 grant 825328 (FASTEN).

\bibliographystyle{IEEEtran}
\bibliography{main}

\end{document}